\documentclass[aps,prl,twocolumn,showpacs,preprintnumbers,amsmath,amssymb]{revtex4}
\usepackage{graphicx}
\usepackage{dcolumn}
\usepackage{bm}
\usepackage[english]{babel}
\usepackage{epsfig}
\usepackage[cp1251]{inputenc}
\usepackage[mathlines]{lineno}
\usepackage{amsmath,amssymb,graphicx,color}

\begin{document}

\title{Resonant transmission of light in arrays of high-index dielectric nanoparticles}

\author{\firstname{Roman~S.} \surname{Savelev}$^1$}
\author{\firstname{Dmitry~S.} \surname{Filonov}$^1$}
\author{\firstname{Mihail~I.} \surname{Petrov}$^1$}
\author{\firstname{Alexander~E.} \surname{Krasnok}$^1$}
\author{\firstname{Pavel~A.} \surname{Belov}$^1$}
\author{\firstname{Yuri ~S.} \surname{Kivshar}$^{1,2}$}
\affiliation{$^1$ITMO University, St.~Petersburg 197101, Russia}
\affiliation{$^2$Nonlinear Physics Centre, Australian National University, Canberra ACT 2601, Australia}

\begin{abstract}
We study numerically, analytically and experimentally the resonant transmission of light in a waveguide formed by a periodic array of high-index dielectric nanoparticles with a side-coupled resonator. We demonstrate that a resonator with high enough $Q$-factor provides the conditions for the Fano-type interference allowing to control the resonant transmission of light. We suggest a practical realization of this resonant effect based on the quadrupole resonance of a dielectric particle and demonstrate it experimentally for ceramic spheres at microwaves.
\end{abstract}

\maketitle

{\em Introduction.} In the past years, a rapid development of the field of all-dielectric nanophotonics is observed aiming at the manipulation of optically-induced electric dipole (ED) and magnetic dipole (MD) resonances in dielectric nanostructures composed of nanoparticles with high refractive index~\cite{EvlyukhinPRB2010,KuznetsovSciRep2012,ReviewSPIE}. Both ED and MD resonant responses were theoretically predicted~\cite{EvlyukhinPRB2010,EvlyukhinPRB2011} and experimentally observed~\cite{KuznetsovSciRep2012,EvlyukhinNL2012} for silicon particles with linear size less than the operating wavelength, and it was shown that the resonance frequencies depend on the particles size and shape. Owing to these properties and to low losses in dielectric materials, different optical nanostructures based on such nanoparticles were suggested~\cite{AhmadiPRA2008,JunjiePRA2009,MiroshnichenkoNL2012,KrasnokUFN2013,CapassoScience2015}. One of such structures is a chain of dielectric nanoparticles with high refractive index. Such structures  were suggested as a realization of optical waveguides with subwavelength light localization and low overall losses required for a design of highly efficient integrated photonic circuits~\cite{JunjiePRA2009,SavelevPRB2014}.

A common practical scheme that employs a coupling between an optical waveguide and a high-$Q$ cavity was suggested for controlling the light propagation and realization an optical bistable devices~\cite{FanOE1998,FanPRB1999,ManolatouJQE1999,YarivPRE2000,FanAPL2002,SoljacicAPL2003,MiroshnichenkoOE2005}. Most of the studies were carried out for photonic-crystal waveguides with different geometries and different types of defects or cavities~\cite{FanOE1998,SoljacicAPL2003,MingaleevPRE2006,MingaleevOE2008,Cun-XiCPB2012,YuAPL2013}. In such a case the interference between the continuum of states (a transmission band of the waveguide) and a discrete resonant state (a resonance of the cavity) provides the conditions for the Fano-type resonance~\cite{MiroshnichenkoRMP2011} realized in a sharp and asymmetric line shape, which reduces the frequency shift and the value of energies required for all-optical nonlinear switching, logic, and information processing~\cite{YangAPL2007,GalliAPL2009}.\par

In this Letter,  we study the resonant transmission of light through a waveguide formed by a chain of high-index dielectric nanoparticles with a side-coupled defect nanoparticle. Localization in such structures in transverse direction is due to the high index of dielectric material and its guiding properties originate from the long-range coupling between the particles. Subwavelength size of the waveguide and the possibility of controlling light transmission makes it promising for designing the efficient optical integrated circuits. The results of this work can be applied to any spectral range, if appropriate materials are provided. Following this idea, we conduct microwave experiments with ceramic spheres and confirm all predicted resonant phenomena.

\begin{figure}[b]\centering
	\includegraphics[width=0.48\textwidth]{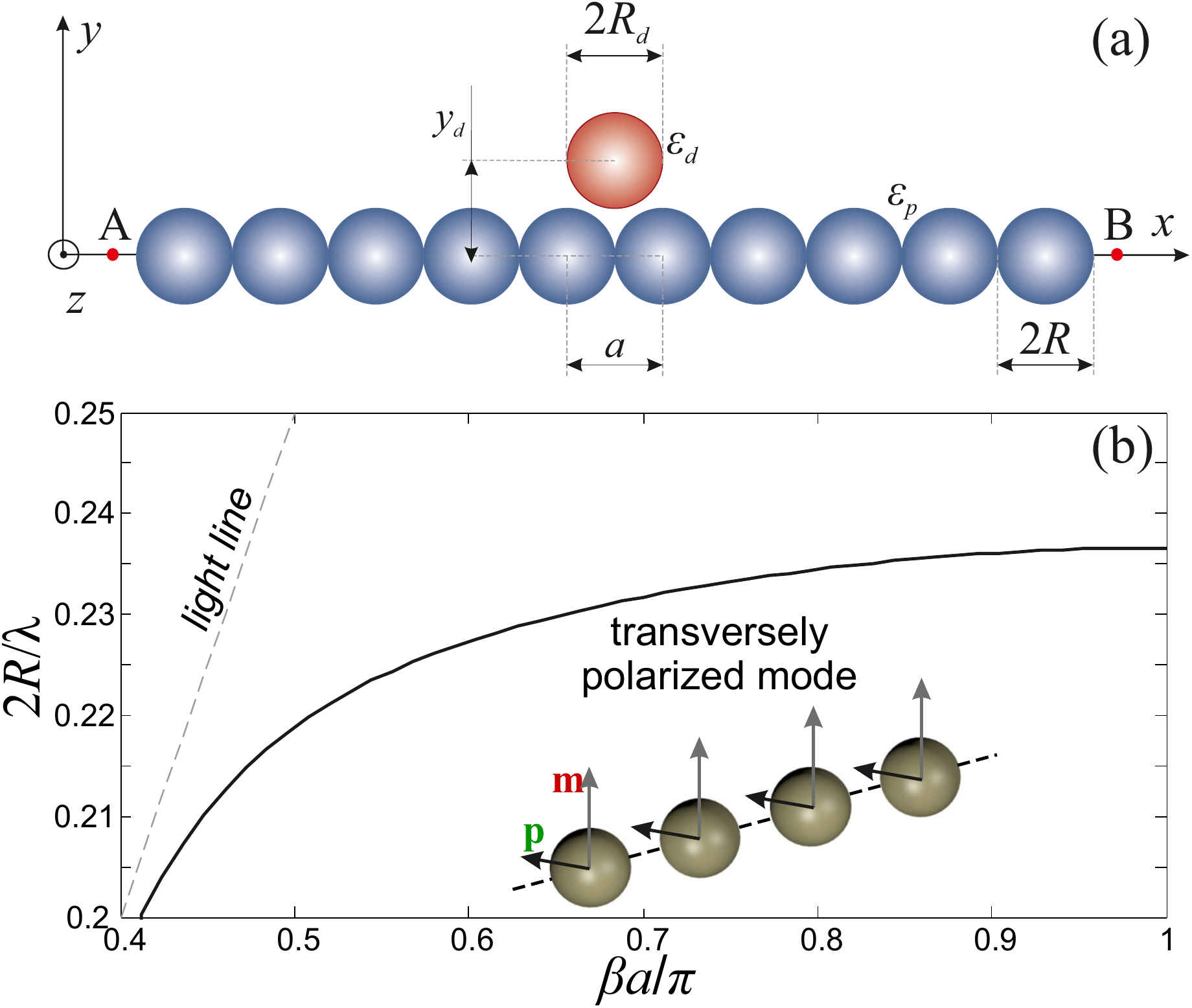}
	\caption{(a) Schematic of a chain of high-index dielectric nanoparticles with permittivity $\varepsilon_p$, radius $R$ and period $a$ coupled to a side particle with permittivity $\varepsilon_d$ and radius $R_d$. ``A'' is a position of the source and ``B'' is the position of the detector. (b) Dispersion curve for the transversely polarized mode of an infinite chain of spherical nanoparticles; $\beta$ is the Bloch wavenumber, $\lambda$ is the wavelength.}
	\label{fig:Scheme}
\end{figure}

{\em Theoretical model.} We consider a chain of closely spaced high-index dielectric nanoparticles with side-coupled defect nanoparticle as shown in Fig.~\ref{fig:Scheme}(a). We fix the permittivity of dielectric particles $\mathrm{Re}(\varepsilon_p)=16$ with small material losses $\mathrm{Im}(\varepsilon_p)=0.02$, that approximately corresponds to silicon nanoparticles in optics~\cite{ZywietzNC2014} and to ceramic spheres at microwaves~\cite{SavelevPRB2014}, and the period of the chain $a=2R$. The radius $R_d$, and the permittivity $\varepsilon_d$ of the side-coupled particle vary for different cases.\par

For theoretical description of the considered system we use analytical approach when applicable, and direct numerical calculations in other cases. Analytical approach is based on the well-known coupled-dipole model, where each particle is modeled as a hybrid of magnetic and electric dipoles with magnetic $\mathbf{m}$ and electric $\mathbf{p}$ momenta, oscillating with frequency $\omega$ [$\propto \exp(-i\omega t)$]. In the CGS system of units such approach formulates as follows:
\begin{align*}
\mathbf{p}_i &={\alpha_e}_i \sum\limits_{j \ne i}\left( \widehat{C}_{ij} \mathbf{p}_j - \widehat{G}_{ij} \mathbf{m}_j \right),\\
\mathbf{m}_i &={\alpha_m}_i \sum\limits_{j \ne i} \left( \widehat{C}_{ij} \mathbf{m}_j + \widehat{G}_{ij} \mathbf{p}_j \right),
\end{align*}
where $\alpha_m$ and $\alpha_e$ are the magnetic and electric polarizabilities of a sphere~\cite{Stratton}, $\widehat{C}_{ij} = A_{ij}\widehat{I} + B_{ij}(\widehat{\mathbf{r}}_{ij} \otimes \widehat{\mathbf{r}}_{ij})$, $\widehat{G}_{ij} = - D_{ij}\widehat{\mathbf{r}}_{ij} \times \widehat{I}$, $\otimes$ is the dyadic product, $\widehat{I}$ is the unit $3 \times 3$ tensor, $\widehat{\mathbf{r}}_{ij}$ is the unit vector in the direction from $i$-th to $j$-th sphere, and
\begin{align*}
A_{ij} &=\dfrac{\exp(i k_h r_{ij})}{r_{ij}} \left( k_h^2-\dfrac{1}{(r_{ij})^2}+\dfrac{i k_h}{r_{ij}} \right),\\
B_{ij} &=\dfrac{\exp(i k_h r_{ij})}{r_{ij}} \left( -k_h^2 + \dfrac{3}{(r_{ij})^2} - \dfrac{3 i k_h}{r_{ij}} \right),\\
D_{ij} &=\dfrac{\exp(i k_h r_{ij})}{r_{ij}} \left( k_h^2 + \dfrac{i k_h}{r_{ij}} \right),
\end {align*}
where $r_{ij}$ is the distance between the centers of $i$-th and $j$-th spheres, $\varepsilon_h $ is the permittivity of the host medium, $k_h=\sqrt{\varepsilon_h}\omega/c$ is the host wavenumber, $\omega=2\pi\nu$, $\nu$ is the frequency, and $c$ is the speed of light. It is known, that such model provides very accurate results for wide range of parameters~\cite{SavelevPRB2014}. In the cases when the dipole model becomes inapplicable, numerical calculations are performed in CST Microwave Studio.\par

\begin{figure}[t]\centering
	\includegraphics[width=1.0\columnwidth]{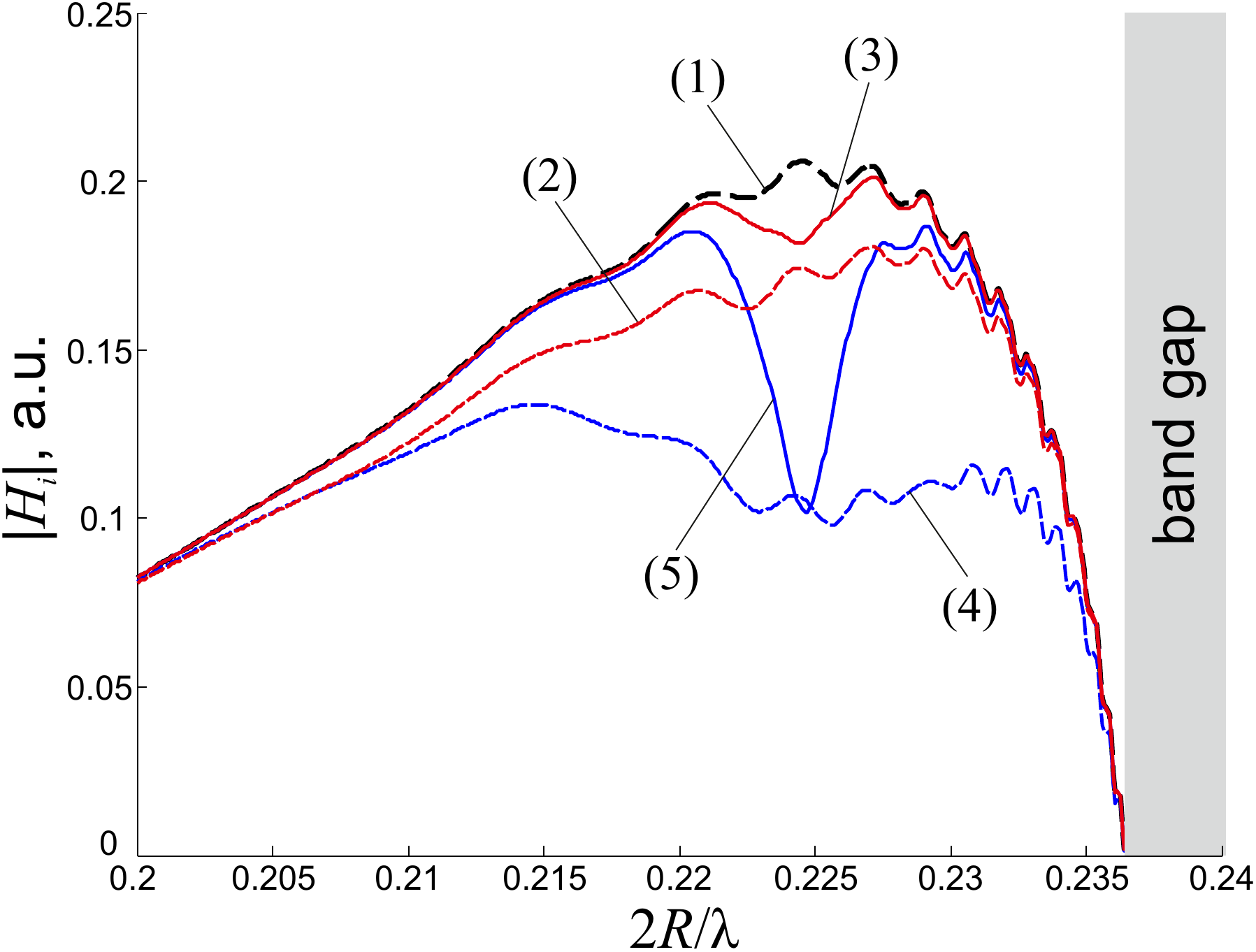}
	\caption{Transmission through the chain of 60 particles for different parameters of the side-coupled sphere: $\varepsilon_d=\varepsilon_p$, $R_d=1.1 R$ [dashed red curve (2) -- $z$-excitation; dashed blue curve (3), $y$-excitation]; $\varepsilon_d=64$, $R_d=0.55 R$ [solid red curve (4) -- $z$-excitation; solid blue curve (5), $y$-excitation]. Black dashed curve (1) corresponds to the case of the chain without side-coupled particle.}
	\label{fig:Transmission}
\end{figure}

{\em Numerical results.} We operate in the frequency range near the magnetic dipole resonance frequency (which is about $2R/\lambda \approx 1/\sqrt{\mathrm{Re}(\varepsilon_p)} = 0.25$) where magnetic moments are dominant. In Fig.~\ref{fig:Scheme}(b) we show the dispersion curves for the transversely (i.e. when magnetic dipoles are oscillating in direction perpendicular to the chain axis) and longitudinally polarized (when magnetic dipoles are oscillating in direction parallel to the chain axis) magnetic modes of the infinite chain with the chosen parameters (assuming $\varepsilon_p$ to be real-valued). Further calculations are carried out for the TM modes.\par

First, we use the side-coupled particle with permittivity $\varepsilon_d=\varepsilon_p=16+0.02i$. The MD resonance frequency of the particles with the chosen parameters lies outside the TM band of the uniform chain [see Fig.~\ref{fig:Scheme}(b)]. By increasing the radius of the side particle to $R_d=1.1R$, we shift its resonance frequency $\omega_d$ to the transmission band. We excite the chain with a point magnetic dipole placed at the distance of $0.5R$ from the left end of the chain [point ``A'' in Fig.~\ref{fig:Scheme}(a)] and calculate the intensity of the $i$-th component of the magnetic field $|H_i|^2$ at the other side of the chain [at the distance $0.5R$ from the right end, point ``B'' in Fig.~\ref{fig:Scheme}(a)]. Which field component is detected is defined by the polarization of the excited chain modes. In this geometry two polarizations can be distinguished: $(m_x,m_y,p_z)$ and $(p_x,p_y,m_z)$. The first type of modes can be excited with the $z$-oriented magnetic dipole and therefore we refer to this case as to the $z$-excitation. We excite the second type of modes with $y$-oriented magnetic dipole ($y$-excitation), since longitudinal $x$-polarized modes lie in different spectral region and we do not consider them here. In this view the case of $z$-excitation is the simplest, since magnetic moments in all particles are oriented in $z$-direction. But in the case of the $y$-excitation magnetic dipoles oscillate in $x-y$ plane and the coupling between the particles is stronger, therefore the stronger effect may be observed.\par

In Fig.~\ref{fig:Transmission} we show $|H_i|^2$ at the point ``B'' for the chain with and without side particle. Here ``$i$'' is $y$ or $z$ for $y-$ or $z$-excitation, respectively. The side particle is located at the distance $y_d=2.5R$ from the chain. In the case of the $z$-excitation (dashed red curve (2) in Fig.~\ref{fig:Transmission}) the transmission remains almost the same, and in the case of the $y$-excitation (dashed blue curve (3) in Fig.~\ref{fig:Transmission}), when the coupling is stronger, we observe a very broad dip that is stretched over the whole pass band. From these calculations it is clear that the $Q$-factor of a single particle is too low to provide some sharp features that can be observed, neither for the $z$-excitation, nor for $y$-excitation. In order to show that side-coupled particle can in fact provide the conditions for Fano-type resonance, we increase its $Q$-factor in the simplest theoretical way -- by increasing its permittivity to $\varepsilon_d=64$. We also decrease the radius of side-coupled particle to $R_d=0.55R$, so its MD resonance frequency remains approximately the same. In this case, as it is shown with solid red curve (4) and blue curve (5) (for $z$- and $y$-excitation, respectively) in Fig.~\ref{fig:Transmission}, we observe the sharp resonance dip in the transmission spectrum of the chain.\par

Further we investigate the case of high-$Q$ side-coupled particle with $\varepsilon_d=64$ in more details. Transmission coefficient through a single-mode waveguide with symmetrically side-coupled resonator, that supports $n$ eigenmodes with the resonance frequencies $\Omega_n$ and the mode decay rate $\Gamma^0_n$ can be calculated with the following formula~\cite{ManolatouJQE1999,YarivPRE2000}:
\begin{equation}
t = 1 - \sum_{n}\dfrac{i\Gamma^c_n}{\Delta\omega_n + i(\Gamma_n^0 + \Gamma_n^c)}; \;\;
T = |t|^2,
 \label{eq:Transmission}
\end{equation}
where $\Delta\omega_n = \omega -  \Omega_n$, and $\Gamma_n^c$ is the decay rate of the $n$-th mode of the side-coupled resonator into the waveguide. Here relative transmission is determined as the ratio of the magnetic field intensities at the detector point $T = |H_i^d|^2/|H_i|^2$ with and without side particle. Note that the side particle also creates the field in the detector point directly through the free space, however this contribution is noticeable only for short chains $\lesssim$ 20 particles. Our calculations are made for the chains consisted of 60 particles, therefore this contribution is negligibly small.\par

\begin{figure}[t]\centering
	\includegraphics[width=1.0\columnwidth]{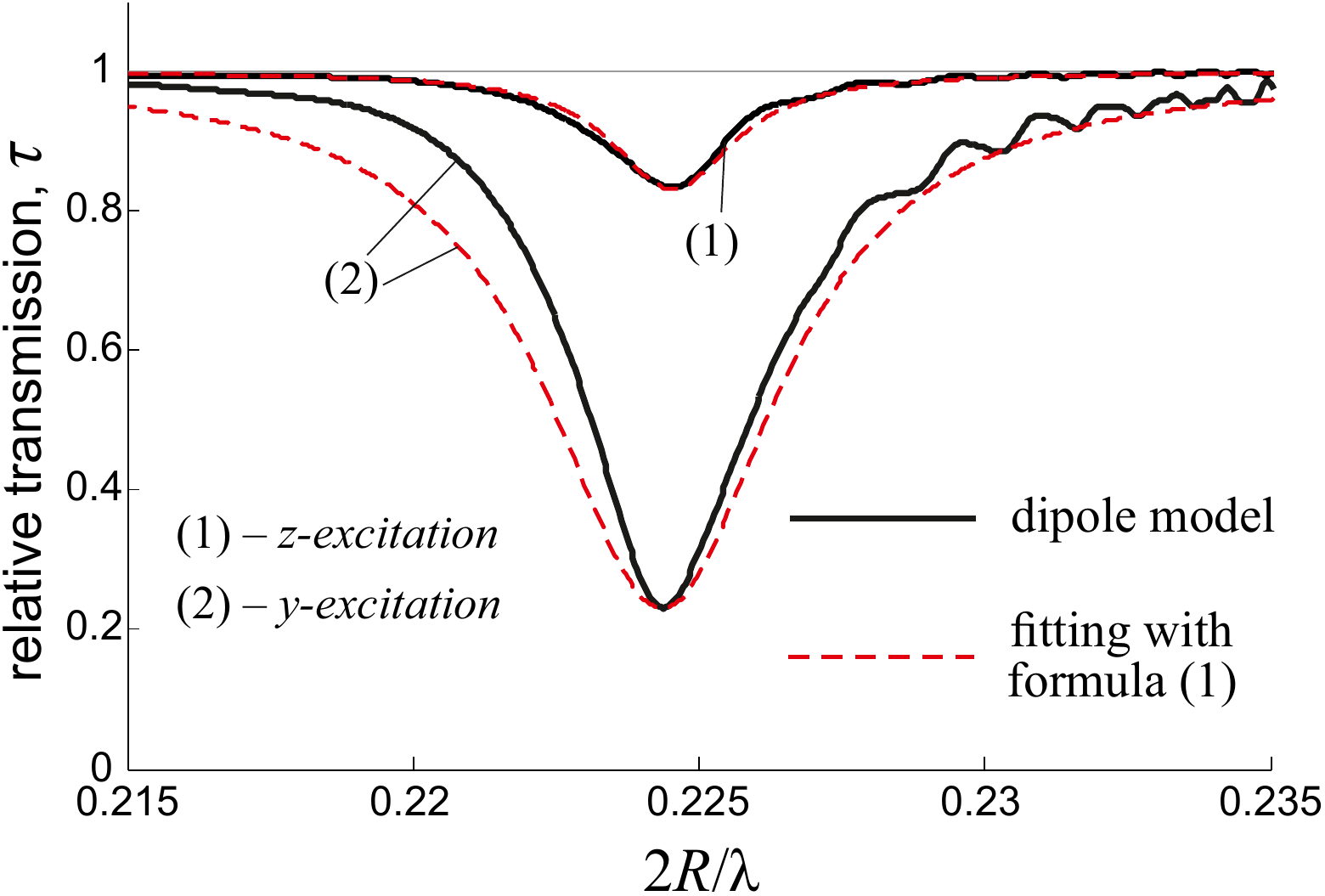}
	\caption{Relative transmission for through a chain with side-located sphere with $\varepsilon_d=64$ and $R_d=0.55 R$; curves (1) correspond to the case $z$-excitation, curves (2) to the case of $y$-excitation. Dashed red curves are obtained via the calculations made within a dipole model and solid black curves via formula~(\ref{eq:Transmission}).}
	\label{fig:Transmission_64}
\end{figure}

In Fig.~\ref{fig:Transmission_64} show the relative transmission $T$ for the cases of $z$- and $y$- excitation. Solid black curves are calculated with dipole model, and dashed red curves with formula~(\ref{eq:Transmission}). In the case of $z$- excitation waveguide mode induces only $z$-oriented magnetic moment on the side-coupled particle, i.e. it interacts only with one MD eigenmode, and transmission coefficient reduces to~\cite{YarivPRE2000}:
\begin{equation}
T = \dfrac{\Delta\omega^2 + (\Gamma^0)^2}{\Delta\omega^2 + (\Gamma^c + \Gamma^0)^2},
 \label{eq:Transmission1}
\end{equation}
In this case almost symmetric dip in transmission is very well described by the formula~(\ref{eq:Transmission1}) for the picked parameter $\Gamma^c$.\par

In the case of $y$- excitation the waveguide mode can propagate via excitation of both $x$- and $y$-oriented magnetic moments on the side-coupled particle. In this case the coupling is stronger and the dip in transmission is deeper. We also observe that the dashed symmetric curve (2) in Fig.~\ref{fig:Transmission_64} fitted with formula~(\ref{eq:Transmission}) can not describe the transmission very accurately, since the solid curve (2) is assymetric. For more accurate desription the dependence of $\Gamma^c$ on the frequency, that causes the assymetry, has to be taken into account. Calculations presented in Fig.~\ref{fig:Transmission_64} confirm that the resonance features in transmission spectra appear as a result of Fano-type interference of the waves that go through resonant and non-resonant channels of transmission.\par

\begin{figure*}[t!]
\includegraphics[width=1.0\textwidth]{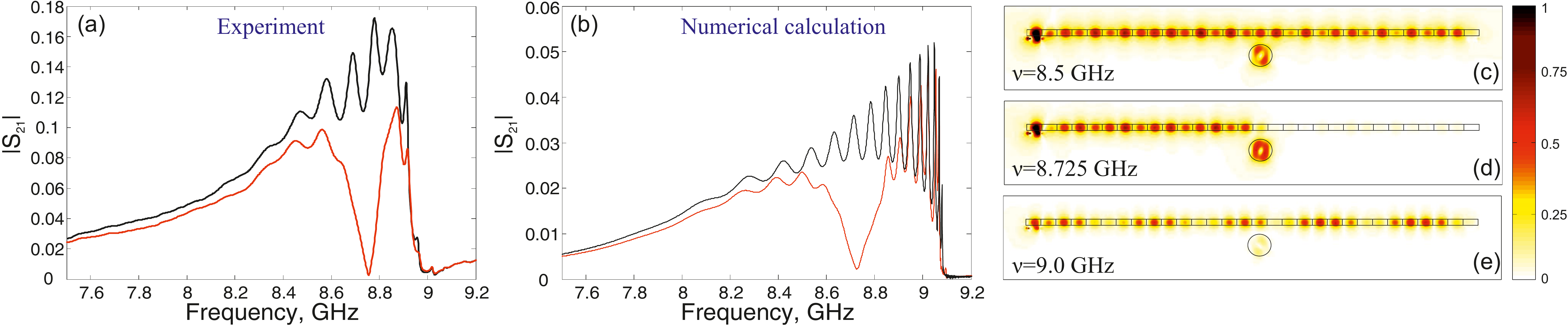}
\caption{(a) Experimentally measured and (b) numerically calculated transmission efficiency for the chains of ceramic disks without (black curve) and with (red curve) side-coupled sphere. (c--e) Magnetic field distribution $|H_z|$ in the chain with side-coupled sphere with $R_d=6.1$\,mm for three frequencies: (b) $\nu=8.5$\,GHz, (c) $\nu=8.725$\,GHz, (d) $\nu=9.0$\,GHz.}
\label{fig:CST}
\end{figure*}


{\em Experimental results.} In order to verify this concept experimentally, we scale the dimensions to the microwave frequency range and perform an experimental study of the transmission efficiency of the waveguide composed of 30 identical ceramic disks with radius $R=3.95$\;mm and height $h=3.9$\;mm and permittivity $\varepsilon \approx 16.4$. The disks are placed in a holder made of a styrofoam material with permittivity close to 1. The axis of the chain is oriented perpendicular to the axis of the disks. The magnetic dipoles are realized as copper loops with the diameter of 2~mm being connected to Agilent E8362C Vector Network Analyzer via coaxial cables. Transmitting and receiving loops are placed at the distance of 1 mm above the first and last disks, respectively.

A practical realization of the resonator with a high value of the $Q$-factor at the subwavelength scale is a complicated task mostly due to radiation losses. As a realistic case, here we suggest to use a sphere with larger radius that exhibits higher-order multipole resonances, with the frequencies tuned to the operational frequency by changing its size. It is known that a dielectric sphere with $\varepsilon \approx 16$ and radius $R$ exhibits a magnetic quadrupole (MQ) resonance at the frequency $\approx 0.175c/R$, with the $Q$-factor being substantially higher than that of MD and ED resonances~\cite{EvlyukhinPRB2011}. For a sphere with $R_d=6.1$\;mm MQ resonance is expected at the frequency $\approx 8.6$\;GHz. In our experiments the permittivity of the spheres was a bit smaller than 16, therefore the MQ resonance frequency is slightly higher.

The magnitude of the experimentally measured transmission coefficient ($|S_{21}|$ parameter) is shown in Fig.~\ref{fig:CST}(a). Black and red curves correspond to a chain without and with side-coupled sphere, respectively. The dip in the transmission (a red curve) is observed at the frequency of the MQ resonance of the side-located sphere. For other frequencies, the transmission efficiency remains approximately the same. We support these experimental results by direct numerical simulations performed with the help of the CST Microwave Studio, since in this case the analytical dipole model is no longer applicable. The calculated transmission for the same parameters as in experiment is presented in Fig.~\ref{fig:CST}(b). We observe very good qualitative agreement between numerical and experimental results. A quantitative difference between the experimentally measured and numerically calculated spectra can be explained by a noticeable dispersion of the size and permittivity of real ceramic disks and by difficulties of an accurate modeling of the transmitting and receiving loops.

In Figs.~\ref{fig:CST}(c-e) we show the calculated magnetic field distribution $|H_z|$ in the plane $y=0$ at three different frequencies. We observe that an additional particle almost completely blocks the propagating wave, which in its turn induces a multipole moment in the particle. Since the considered system is open,  a weak scattering of radiation is also observed.

In summary, we have demonstrated theoretically, numerically and experimentally the possibility of an resonant control of the light transmission through waveguides composed of high-index dielectric nanoparticles with side-coupled resonators. We have demonstrated that an effective control requires a resonator with high enough $Q$-factor, that can be realized for a dielectric sphere with a larger radius
supporting high-$Q$ multipole resonance in the operating spectral range.

{\em Acknowledgements.} This work was supported by the Government of Russian Federation (projects GZ 2014/190, GZ 3.561.2014/K, Grant 074-U01), the Russian Foundation for Basic Research (project No.~14-02-31761), the Dynasty Foundation (Russia), and the Australian Research Council.

\end {document}